\documentclass{ieee} 
\usepackage{times}
\usepackage{makeidx}  
\usepackage{graphicx}
\usepackage{amssymb}

\newtheorem{definition}{Definition}
\newtheorem{theorem}{Theorem}
\newtheorem{lemma}{Lemma}
\begin{document}

\title{Scalable Probabilistic Models for 802.11 Protocol Verification}
\author{Amitabha Roy and K. Gopinath\\
Computer Architecture and Systems Laboratory\\ 
Department of Computer Science and Automation \\ Indian Institute of Science, Bangalore\\
\{aroy,gopi\}@csa.iisc.ernet.in}

\maketitle
\thispagestyle{empty}

\begin{abstract}
The IEEE 802.11 protocol is a popular standard for wireless local area
networks. Its medium access control layer (MAC) is a carrier sense
multiple access with collision avoidance (CSMA/CA) design and includes
an exponential backoff mechanism that makes it a possible target for
probabilistic model checking.  In this work, we identify ways to
increase the scope of application of probabilistic model checking to
the 802.11 MAC. Current techniques do not scale to networks of even
moderate size.  To work around this problem, we identify properties of
the protocol that can be used to simplify the models and make
verification feasible. Using these observations, we directly optimize
the probabilistic timed automata models while preserving probabilistic
reachability measures. We substantiate our claims of significant
reduction by our results from using the probabilistic model checker PRISM.
\end{abstract}
\section{Introduction}
The IEEE 802.11 protocol \cite{standard} is a popular standard for wireless networks. Its
medium access control layer (MAC) is a carrier sense multiple access
with collision avoidance (CSMA/CA) design and includes an exponential
backoff mechanism that makes it an ideal target for probabilistic
model checking.  This protocol has been modeled
using a range of techniques such as finite state machines \cite{fsm} and
probabilistic timed automata \cite{wlan}.

The 802.11 protocol suffers from a potential livelock problem,
demonstrated formally in \cite{fsm}, which is mitigated only by the
presence of a finite retry limit for each data packet. The livelock
arises because it is possible, although improbable, for two stations
to behave symmetrically and continuously collide until they drop their
respective packets on exceeding the retry limit. In such a scenario, it
is useful to bound the probability of such pathologically symmetric behavior. 
This motivates the application of probabilistic model checking to the
problem of computing probabilities of desired and undesired
behavior in the protocol. Two primary properties of interest are: the probability of
the number of retries reaching a certain count and the probability of
meeting a \emph{soft} deadline.

A recent solution to the problem of obtaining these probabilities 
has been proposed in \cite{wlan}. It
models a limited (but critical) aspect of the protocol using
Probabilistic Timed Automata (PTA) \cite{pta} and exploits available
tools, namely, the Probabilistic Symbolic Model Checker (PRISM)
\cite{prismpub,website} for computing the probability values and the real time
model checker Uppaal \cite{uppaal} as a proof assistant. Results on
the probability of the backoff counter on a station reaching a
particular value and the probability of a packet being transmitted
within a certain deadline are presented. This work, however, suffers
from scalability problems.  The model assumes only two stations
(sender destination pairs). When we extended the models to 3
stations (and 3 corresponding destinations), which is a
practical sized network topology, we found it computationally
infeasible to model check properties of interest. These problems are
compounded by an inaccurate assumption that the packet length can vary
on every retransmission.

The aim of this work is twofold. First, we present a more accurate and scalable model
for the protocol. Second, we set up a logical framework to exploit protocol specific
redundancies. Under this framework, we perform a
number of provably correct optimizations that reduce the generalized multi station model. 
The optimizations involve abstracting away the deterministic waits, and considering only a 
subset of the allowed packet sizes that nevertheless captures all the
relevant behavior. In addition, we duplicate the model reduction technique of 
\cite{wlan} for the multi station problem. 

Our reduced models are immediately verifiable in PRISM and require no
further tools. However, the option of using tools like RAPTURE
\cite{rrs} on the reduced PTA models remains. Our results show a
reduction in state space over the existing solution for 
two stations. We are also able to successfully model check 
a topology of three station that was infeasible with the current models.

The organization of the paper is as follows. We begin with the  
modeling formalism used in this paper. We present the scalable models 
for the  multi station 802.11 problem and discuss the behavior of the
protocol. Next, 
we present a notion of equivalence in probabilistic systems that abstracts away 
deterministic deterministic paths in the system but preserves probabilistic reachability. 
We give sufficient requirements for equivalence both at the level of untimed probabilistic systems and
probabilistic timed automata. Based on this framework, we present our set of reductions to the
generalized model for the multi station problem. We also show that 
we can verify soft deadlines inspite of these optimizations. We conclude with 
results that detail state space reduction as well as case studies for a three station topology.

\section{Modeling formalism for the 802.11 protocol}
\label{sect:formalism}
In order to efficiently model and verify the 802.11 protocol, we need
a modeling formalism that can represent the protocol at
sufficient depth and, at the same time, must be amenable to
transformations for more efficient verification. We have been guided by the
existing work in \cite{wlan} in our choice of Probabilistic Timed
Automata to model the 802.11 protocol.  

We introduce Probabilistic Timed Automata (PTA) \cite{pta}, Probabilistic
Systems (PS) \cite{wlan,psim} and fully probabilistic systems (FPS). All these have
been surveyed in \cite{theorypct} with special reference to their
relationship in the context of probabilistic model checking.

Let $\chi$ be a set of non-negative real valued variables called
clocks. Call $Z$ the set of zones over $\chi$, which is the set of all
possible atomic constraints of the form $x \sim c$ and $(x-y)\sim c$
and their closure under conjunction. Here $x,y\in \chi$, $\sim$
$\in$ $\{<,\leq,>,\geq\}$ and $c \in \mathbb{N}$, where $\mathbb{N}$ is the set of natural numbers.
A clock valuation
$v$ is the assignment of values in $\mathbb{R}_{\geq 0}$(where
$\mathbb{R}_{\geq 0}$ is the set of non-negative reals) to all
clocks in $\chi$.  The concept of a clock valuation $v$ satisfying a
zone $Y$, indicated as $v\triangleleft Y$, is naturally derived by
assigning values to each clock in the zone and checking whether all
constraints are satisfied.

\begin{definition}
A probabilistic timed automaton is a tuple $(L,\overline{l},\chi,\Sigma,I,P)$ where 
$L$ is a finite set of states, $\overline{l}$ is the initial state, $\chi$ is the set of
clocks and $\Sigma$ is a finite set of labels used to label transitions.
The function $I$ is a map $I:L\rightarrow Z$ called the invariant
condition. The probabilistic edge relation $P$ is defined as
$P\subseteq L \times Z \times \Sigma \times \mathit{Dist}(2^\chi \times L )$,
where $\mathit{Dist}(2^\chi \times L )$ is the set of all probability distributions, each
elementary outcome of which corresponds to resetting some clocks to
zero and moving to a state in $L$. We call a distinguished (not necessarily non-null)
subset $\Sigma^u$ of the set of events as \emph{urgent events}.
\end{definition}

A critical feature of PTAs that makes them powerful modeling tools is
that each transition presents \emph{probabilistic choice} in the PTA
while different outgoing probabilistic transitions from a state
present \emph{non-deterministic} choice in the PTA. Hence, a PTA can
model non-determinism, which is inherent in the composition of
asynchronous parallel systems.

Composition of PTAs is a cross product of states with the condition that 
the composed PTAs must synchronize on shared actions. For a
detailed description see \cite{wlan}.

A feature of PTAs that is useful for higher-level modeling is urgent
channels. Urgent channels are a special set of edge labels (symbols)
on which a PTA must synchronize whenever possible.

\begin{definition}
A probabilistic system (PS), is a tuple $(S,\overline{s},\Sigma,\mathit{Steps})$
where $S$ is a finite set of states, $\overline{s}$ is the start state, $\Sigma$ is a
finite set of labels and $\mathit{Steps}$ is a function $\mathit{Steps}:S \rightarrow 2^{\Sigma \times
\mathit{Dist}(S)}$ where $\mathit{Dist}(S)$ is the set of all distributions over $S$.
\end{definition}
This is the same as the simple probabilistic automaton of \cite{psim}.
\begin{definition} \label{pta2psdef}
Given a PTA $\mathcal{T}$$=(L,\overline{l},\chi,\Sigma,I,P)$, the \emph{semantics} of 
$\mathcal{T}$ is the Probabilistic System 
$[[\mathcal{T}]]$$=(S,\overline{s},\mathit{Act},\mathit{Steps})$, 
with the following definitions:\\ 
$S \subseteq L \times {\mathbb{R}}_{\geq 0}^{\mid\chi\mid}$ is 
the set of states  with the restrictions 
$(s,v) \in S$ iff ($s \in L$ and $v \triangleleft I(l)$) and 
$\overline{s}=(\overline{l},0)$.

$\mathit{Act}={\mathbb{R}}_{\geq 0} \cup \Sigma$. This reflects either actions 
corresponding to time steps (${\mathbb{R}}_{\geq 0}$) or actions from the PTA ($\Sigma$). 

$\mathit{Steps}$ is the least set of probabilistic transitions containing, 
for each $(l,v) \in S$, a set of action distribution pairs $(\sigma, \mu)$ 
where $\sigma$ $\in$ $\Sigma$ and $\mu$ is a probability 
distribution over $S$. $\mathit{Steps}$ for a state $s=(l,v)$ is defined as follows.\\
I. for each $t \in \mathbb{R}_{\geq 0}$ $(t,\mu) \in \mathit{Steps}(s)$ iff
\begin{enumerate}
\item  $\mu(l,v+t)=1$ and $v+t' \triangleleft I(l)$ for all $0 \leq t' \leq t$.
\item For every probabilistic edge of the form $(l,g,\sigma,-) \in P$, if $v+t' \triangleleft g$ for any 
$0 \leq t' \leq t$, then $\sigma$ is non-urgent.
\end{enumerate}
II. for each $(l,g,\sigma,p) \in P$, let $(\sigma,\mu)\in \mathit{Steps}(s)$ iff 
$v \triangleleft g$ and for each $(l',v') \in S$:
$\mu(l',v')=\Sigma_{X \subseteq \chi \& v'=v[X:=0]}$ $p(X,l')$, the sum being over 
all clock resets that result in the valuation $v'$.
\end{definition}
A critical result \cite{integer}, analogous to the region construction result for 
timed automata, states that it is sufficient to assume only integer increments 
when all zones are closed (there are no strict inequalities). Hence, the definition given above
is modified to $S \subseteq L \times {\mathbb{N}}^{\mid\chi\mid}$ and 
$Act={\mathbb{N}} \cup \Sigma$. Under integer semantics, the size of 
the state space is proportional to the largest constant used. For the rest of this paper, 
we will assume integer semantics. 

Note that, in the presence of non-determinism, the probability measure of a path in a PS is undefined.
Hence, define an adversary or scheduler that resolves non-determinism as follows:
\begin{definition}
An adversary of the Probabilistic System $\mathcal{P}=(S,\overline{s},\mathit{Act},\mathit{Steps})$ 
is a function $f:S \rightarrow \cup_{s \in S} \mathit{Steps}(s)$ where $f(s) \in \mathit{Steps}(s)$.
\end{definition}

We only consider \emph{simple} adversaries that do not change their decision about an outgoing
distribution every time a state is revisited, their sufficiency has been shown in \cite{simple}. 
A simple adversary induces a Fully Probabilistic System (FPS) as defined below.
\begin{definition}
A simple adversary $A$ of a Probabilistic System $\mathcal{P}=(S,\overline{s},\mathit{Act},
\mathit{Steps})$ induces a Fully Probabilistic 
System (FPS) or Discrete Time Markov Chain $\mathcal{P}^A=(S,\overline{s},P)$. Here,
$P(s)=A(s)$, the unique outgoing probability distribution for each $s \in S$, where we drop
the edge label on the transition.
\end{definition}

Thus, given a PS $\mathcal{M}$ and a set of ``target states'' $F$, consider an adversary $A$ and the 
corresponding FPS $\mathcal{M}^A$. A probability space ($\mathit{Prob}^A$) may be defined on 
$\mathcal{M}^A$ via a cylinder construction \cite{cylinder}. A path $\omega$ in $\mathcal{M}^A$ 
is simply a (possibly infinite) sequence of states $\overline{s}s_1s_2...$ such that there is a 
transition of non-zero probability between any two consecutive states in the path. For model 
checking, we are interested in \\ 
$\mathit{ProbReach}^A(F)\stackrel{def}{=}\mathit{Prob}^A\{\omega \in \mathit{Path}^A_{\infty}\mid\exists i\in \mathbb{N} 
\mbox{ where }\omega(i) \in F\}$. $F$ is the desired set of target states, $\omega(i)$ is the 
$i^{th}$ state in the path $\omega$ and $\mathit{Path}^A_{\infty}$ represents all infinite paths 
in $\mathcal{M}^A$. 
Define $\mathit{MaxProbReach}^M(F)$ and $\mathit{MinProbReach}^M(F)$ as the supremum and infimum 
respectively of $\{\mathit{ProbReach}^A(F)\}$ where the quantification is over all adversaries.

\section{Logic Formulas Under Consideration}

Properties of interest at the PTA level are specified using
Probabilistic Computational Tree Logic (PCTL) formulas \cite{hans}.
We limit ourselves to restricted syntax (but non trivial) PCTL
formulas, expressible as $P_{\sim \lambda}\{\Diamond p\}$, where
$\sim \in \{<,>,\leq,\geq\}$ and $\lambda$ is the constant probability
bound that is being model checked for. These PCTL formulas translate
directly into a probabilistic reachability problem on the semantic
Probabilistic System corresponding to the PTA. The reason for this
restriction is that, in the case of the 802.11 protocol, the properties
of interest, including the real time ones, are all expressible in this form. 
For example, in the case of
a probabilistic timed automaton $\mathcal{A}$, the PCTL formula
$P_{<0.5}$ $\{\Diamond p\}$ directly translates to maximum
probabilistic reachability on the induced Markov decision process
$[[\mathcal{A}]]$ from a well-defined start state. We mark the target
states as those where the proposition $p$ is true. The model checker
returns true when this maximum probability is smaller than 0.5.  Under
this restricted form of PCTL, we indicate numerical equivalence using
the following notation.
\begin{definition}
Two probabilistic systems $\mathcal{P}_1$ and $\mathcal{P}_2$ 
are equivalent under probabilistic reachability of 
their respective target states $F_1$ and $F_2$, denoted by \\ 
$\mathcal{P}_1 \stackrel{\mathit{PS}}{\equiv}_{F_1,F_2} \mathcal{P}_2$ when
$\mathit{MaxProbReach}^{\mathcal{P}_1}(F_1)=\mathit{MaxProbReach}^{\mathcal{P}_2}(F_2)$\\
and $\mathit{MinProbReach}^{\mathcal{P}_1}(F_1)=\mathit{MinProbReach}^{\mathcal{P}_2}(F_2)$.
\end{definition}
\begin{definition}\label{def:eqpta}
$\mathit{PTA}_1 \stackrel{\mathit{PTA}}{\equiv}_{\phi_1,\phi_2} \mathit{PTA}_2$ when 
$[[\mathit{PTA}_1]] \stackrel{\mathit{PS}}{\equiv}_{F_1,F_2} [[\mathit{PTA}_2]]$.
The criterion for marking target states is that $F_1$ corresponds to the target states in the
reachability problem for the PCTL formula $\phi_1$, while $F_2$ corresponds to the target states for 
the PCTL formula $\phi_2$.
\end{definition}
\section{Probabilistic Models of the 802.11 Protocol}
\label{sect:probmod}
\begin{figure}[h!]
\includegraphics[width=8cm,height=10cm]{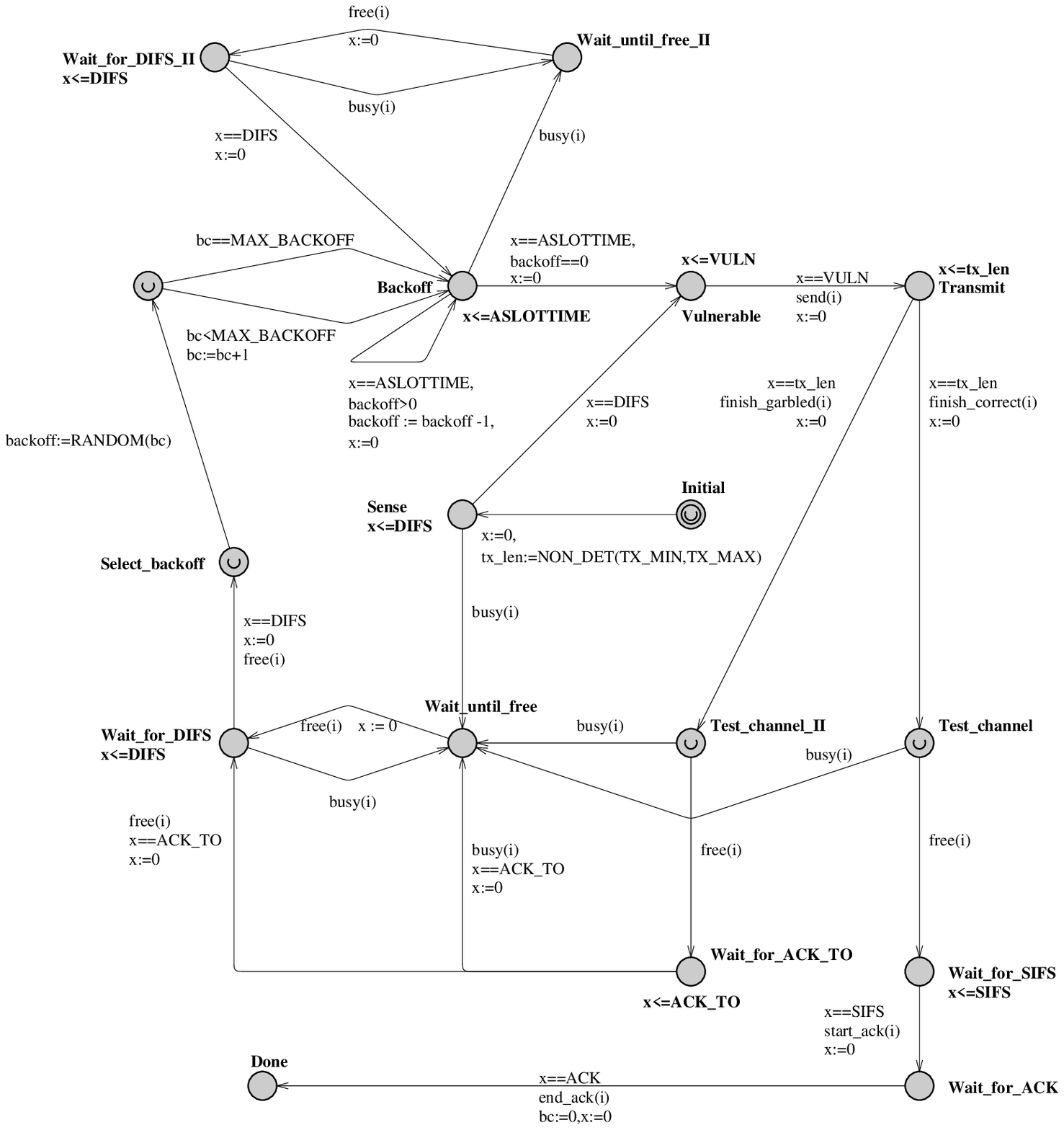}
\caption{PTA model for an Abstract Station - represents both the sender and destination}
\label{fig:abstn}
\end{figure}

\begin{figure}[h]
\includegraphics[width=8cm,height=6cm]{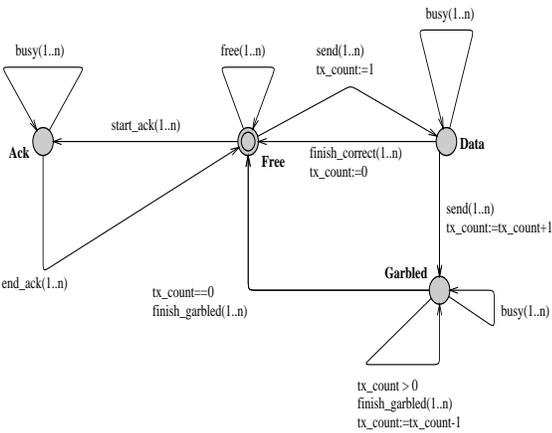}
\label{fig:roychn}
\caption{PTA model for the Channel - Generalized for the multiple station case}
\end{figure}

In this section, we present scalable probabilistic models of the
802.11 basic access MAC protocol assuming no hidden nodes\footnote{In
the absence of hidden nodes \cite{hidden}, the channel is a shared medium visible to
all the stations.}. The model for the \emph{multi-station} 802.11
problem consists of the station model and a shared channel,
shown in Figures \ref{fig:abstn} and \ref{fig:roychn}
respectively. We assume familiarity with
conventions used in graphical representation of timed automata. In
particular, the states marked with a 'u' are urgent states while that
marked by concentric circles is the start state. The station models are
are replicated to represent multiple
sender-destination pairs. Some critical state variables are: $\mathit{bc}$ that
holds the current backoff counter value, $\mathit{tx\_len}$ that holds the
chosen transmission length and \emph{backoff} that represents the
current remaining time in backoff.  The function $\mathit{RANDOM}(\mathit{bc})$ is a
modeling abstraction that assigns a random number in the current
contention window. Similarly, $\mathit{NON\_DET}(\mathit{TX\_MIN},\mathit{TX\_MAX})$ assigns a
non-deterministic packet length between $\mathit{TX\_MIN}$ and $\mathit{TX\_MAX}$, which
are the minimum and maximum allowable packet transmission times
respectively. The values used for verification are from the Frequency
Hopping Spread Spectrum (FHSS) physical layer \cite{standard}. The
transmission rate for the data payload is 2 Mbps.

The station automaton shown in Figure \ref{fig:abstn}, begins with a data
packet whose transmission time it selects non-deterministically in
the range from $258 \mu s$ to $15750 \mu s$. On sensing the channel free
for a Distributed InterFrame Space $(\mathit{DIFS}=128 \mu s)$, it enters 
the $\mathit{Vulnerable}$ state, where it
switches its transceiver to transmit mode and begins transmitting the
signal. The $\mathit{Vulnerable}$ state also accounts for propagation delay. It
moves to the $\mathit{Transmit}$ state after a time $\mathit{VULN}=48 \mu s$ with a
synchronization on $\mathit{send}$. After completing transmission, the station
moves to $\mathit{Test\_channel}$ via one of the two synchronizations,
$\mathit{finish\_correct}$ on a successful transmission and $\mathit{finish\_garbled}$ on an
unsuccessful transmission. The channel keeps track of the status of
transmissions, going into a garbled state whenever more than one
transmission occurs simultaneously. The station incorporates the behavior of
the destination and diverges depending on whether the transmission 
was successful, or not. If the transmission was successful, the portion of the station
corresponding to the destination waits for a Short InterFrame Space $(\mathit{SIFS}=28 \mu s)$ 
amount of time before transmitting an ack, which takes $\mathit{ACK}=183 \mu s $ amount of
time.

On an unsuccessful transmission, the station waits for the
acknowledgment timeout of $\mathit{ACK\_TO}=300 \mu s $. It then 
enters a backoff phase, where it probabilistically selects a random
backoff period \emph{backoff}$=\mathit{RANDOM}(\mathit{bc})$.  $\mathit{RANDOM}(\mathit{bc})$ 
is a function that selects with uniform probability, a value from the contention window
given by the range $[0, (C +1). 2^{\mathit{bc}} -1 ]$, where $C$ is the
minimum contention window ($15 \mu s$ for the FHSS physical
layer). The backoff counter ($\mathit{bc}$) is incremented each time the station
enters backoff. The backoff counter is frozen when a station detects a
transmission on the medium while in backoff.

The station and channel models are different from those in
\cite{wlan}. The station now fixes a packet transmission length
non-deterministically and remembers it rather than allow it to vary on
every retransmission. The channel of \cite{wlan} assumes a fixed topology of
two stations, while the channel depicted in Figure \ref{fig:roychn} is
generalized for an arbitrary number of stations. It follows a
different design from that in \cite{wlan}, which if generalized would
have states exponential in the number of stations. Ours is only
linear. Since the models are generalized to an arbitrary number of
stations, the synchronization labels have subscripts
indicating the station number. However, in the rest of the paper we
drop subscripts whenever the station number is clear from the context.

We point out here that we start with an abstracted station model, which incorporates
the deterministic destination. That this is a valid abstraction has already been shown
for the two station case in \cite{wlan}. The extension to the multi station case 
does not represent any significant new result and hence has been omitted.

\section{Compression of Deterministic Paths: A Technique for State Space Reduction}
\label{sect:red}
\label{subsect:cdp}
In the 802.11 protocol, there are numerous cases where the component
automata representing the system simply count time or where different
resolutions of non-determinism lead to same state but through different paths. 
If we are verifying an untimed property then such fine grained analysis increases state space
without any contribution to probabilistic reachability. We discovered on 
studying these models that it is possible to derive alternative \emph{optimized} 
probabilistic timed automata that avoid the cost of such unnecessary deterministic behavior by
compressing these deterministic paths into equivalent but shorter
paths. The problem is the lack of a suitable formalism to support
our optimizations. This section provides a framework that can be used
to justify the equivalence of our optimized models to the original
ones.

For purposes of comparison, we assume that the state space is a subset of an 
implicit global set of states.  This allows operations such as
intersection and union between the set of states of two different
automata. In particular, for this paper we consistently name states
across the automata we consider.

Our objective is to formalize ``deterministic" behavior of interest.
The key relationship used in this formalization is a specialization 
of dominators as defined in \cite{rrs}.  We refer to this restricted version of dominators as
``deterministic dominators" in the rest of this paper.

\begin{definition}
For a distribution $\pi$ over the \emph{finite} elementary event set $X$, define the support of the 
distribution as $\mathit{supp}(\pi)= \{ x \in X  \mid \pi(x) > 0\} $  
\end{definition}
\begin{definition}
Given a probabilistic system consisting of the set of states $S$,
define $ \prec_D$ as the smallest relation in $S \times S$ satisfying the following:
$\forall s \in S$\\$s \prec_D s$ and \\ $\exists t \in S\mbox{\space} [\forall (a,\pi) \in \mathit{Steps}(s): \exists x\mbox{\space\space\space}(\mathit{supp}(\pi)=\{x\})$ $\wedge$ $(x \prec_D t)]$ $\Rightarrow$ $s \prec_D t$
\end{definition}
If the relation $s \prec_D t$ holds then we say that $t$ is the deterministic dominator of 
$s$. 

An example of a deterministic dominator is shown in the probabilistic systems of Figure 
\ref{fig:ps}, where $S \prec_D T$. 
\begin{definition}
Given distributions $P_1$ over $S_1$ and $P_2$ over $S_2$, define $P_1 \stackrel{\mathit{dist}}{\equiv}P_2$
when $\mathit{supp}(P_1)=\mathit{supp}(P_2)=S$ and $\forall s \in S$ we have $P_1(s)=P_2(s)$.
\end{definition}
Based on the notion of equivalence of distributions, we 
define the notion of equivalence of \emph{sets} of distributions.
Let $\mathit{Steps}_1$ be a set of labeled distributions over $S_1$ and $\mathit{Steps}_2$ 
be a set of labeled distributions over $S_2$.
\begin{definition}
$\mathit{Steps}_1\stackrel{\mathit{dist}}{\equiv}\mathit{Steps}_2$ whenever $\forall (a,\mu_1) \in \mathit{Steps}_1$
$ \exists (b,\mu_2) \in \mathit{Steps}_2$ such that
$\mu_1 \stackrel{\mathit{dist}}{\equiv} \mu_2$ and $\forall (a,\mu_2) \in \mathit{Steps}_2$ $ \exists (b,\mu_1) \in \mathit{Steps}_1$ with
$ \mu_2 \stackrel{\mathit{dist}}{\equiv}\mu_1$.
\end{definition}
Define a path in a probabilistic system as follows:
\begin{definition}
A path in the probabilistic system $\mathcal{P}=(S,\overline{s},\Sigma,\mathit{Steps})$ is a sequence
of state-action pairs $(s_1,a_1),(s_2,a_2)..(s_{n+1})$ such that $\forall i \in \{1..n\}$ we have
$\exists (a_i,\mu)\in \mathit{Steps}(s_i)$ such that $\mu(s_{i+1})>0$.
\end{definition}
\subsection{Deterministic Path Compression in Probabilistic Systems}
\begin{figure}[t]
\includegraphics[width=7cm]{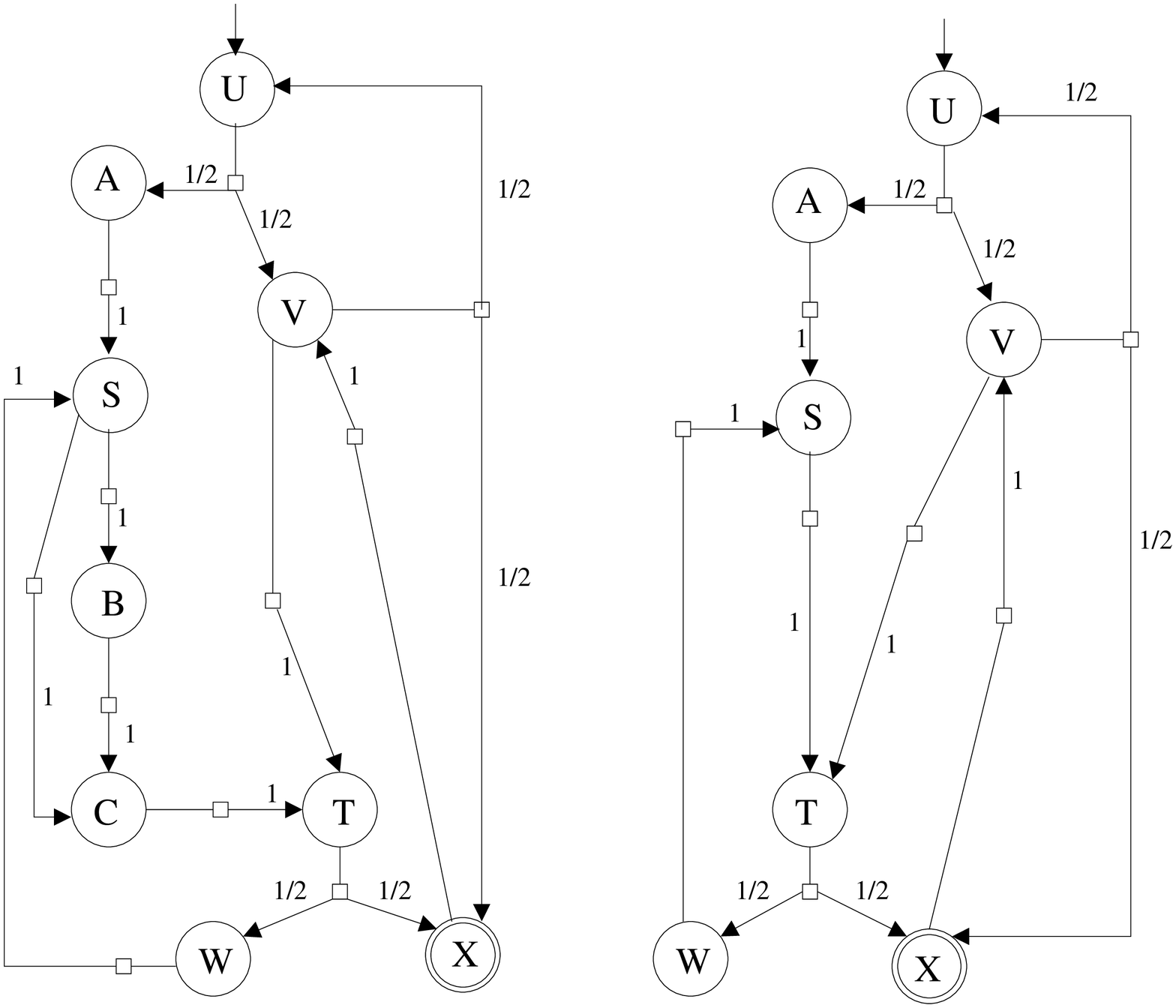}
\caption{Two related Probabilistic Systems}
\label{fig:ps}
\end{figure}

Consider the two probabilistic systems of Figure \ref{fig:ps}, each of which has the
start state $U$.
It should be clear that each of $\mathit{MaxProbReach}({X})$ and $\mathit{MinProbReach}({X})$
takes the same value in both the systems since we have only removed (compressed) the deterministic 
segment $B\rightarrow C$.
We formalize this notion of deterministic path compression 
at the level of probabilistic systems in theorem \ref{thm:ps}.
 
Consider two \emph{finite} probabilistic systems 
$\mathit{PS}_1(S_1,\overline{s},\mathit{Act},\mathit{Steps}_1)$ and 
$\mathit{PS}_2(S_2,\overline{s},\mathit{Act},\mathit{Steps}_2)$ with an identical set of actions. 
All transitions in $\mathit{Steps}_1$ and $\mathit{Steps}_2$ are simple 
transitions of the form $(s,a,\mu)$ where $s$ 
is the originating state, $a\in \mathit{Act}$ and $\mu$ is a probability distribution over 
the state space. Note that the $S_1$ and
$S_2$ are necessarily not disjoint because of the common start state $s$.
\begin{definition}
If, for some $s \in S_1 \cap S_2$, $\mathit{Steps}_1(s) \stackrel{\mathit{dist}}{\equiv} \mathit{Steps}_2(s)$ does not 
hold then $s$ is a point of disagreement between the two probabilistic systems.
\end{definition}
\begin{theorem}[Equivalence in Probabilistic Systems] \label{thm:ps} 
Given two probabilistic systems $\mathit{PS}_1(S_1,\overline{s},\mathit{Act},\mathit{Steps}_1)$ and $\mathit{PS}_2(S_2,\overline{s},\mathit{Act},\mathit{Steps}_2)$
satisfying the following conditions:
\begin{enumerate}
\item For any state $s \in S_1 \cap S_2$, if
$s$ is a point of disagreement then
$\exists t \in S_1 \cap S_2$ such that, $t$ is not a point of disagreement and in each of the systems, $s \prec_D t$. 
\item 
Let $F_1 \subseteq S_1$ and $F_2 \subseteq S_2$ be sets of target states we are model checking for. We impose 
the condition $S_1 \cap S_2 \cap F_1=S_1 \cap S_2 \cap F_2$. For every $s \in S_1 \cap S_2$, which is
a point of disagreement we have the following:  
For the postulated deterministic dominator $t$ and for every state $u$ on any 
path in $\mathit{PS}_1$ between $s$ and $t$, 
$u \in F_1  \Rightarrow (s \in F_1) \vee (t \in F_1)$. Similarly, for every state $u$ on any 
path in $\mathit{PS}_2$ between $s$ and $t$, $u \in F_2 \Rightarrow (s \in F_2) \vee (t \in F_2)$. 
\end{enumerate}
Under these conditions, $\mathit{PS}_1\stackrel{\mathit{PS}}{\equiv}_{F_1,F_2}\mathit{PS}_2$.
\end{theorem}
The proof follows from first principles by setting up a bijective mapping between 
paths in the two probabilistic systems. The complete proof is available in \cite{full}.
\subsection{Equivalence of Probabilistic Timed Automata}
Given two Probabilistic Timed Automata $\mathit{PTA}_1$ and $\mathit{PTA}_2$ and their respective
restricted PCTL requirements 
$\phi_1$ and $\phi_2$, we need a set of  
conditions under which we may claim $\mathit{PTA}_1 \stackrel{\mathit{PTA}}{\equiv}_{\phi_1,\phi_2} 
\mathit{PTA}_2$.
By Definition \ref{def:eqpta}, this is equivalent to showing that 
$[[\mathit{PTA}_1]] \stackrel{\mathit{PS}}{\equiv}_{F_1,F_2} [[\mathit{PTA}_2]]$, where $F_1$ and $F_2$ are the 
corresponding target states of $\phi_1$ and $\phi_2$ respectively.
Our optimizations are based on deterministic path compression as outlined in Section
\ref{subsect:cdp}. Hence, we impose requirements on $\mathit{PTA}_1$ and $\mathit{PTA}_2$ under which
we can apply theorem \ref{thm:ps} to $[[\mathit{PTA}_1]]$ and $[[\mathit{PTA}_2]]$ to deduce 
$[[\mathit{PTA}_1]] \stackrel{\mathit{PS}}{\equiv}_{F_1,F_2} [[\mathit{PTA}_2]]$. The following lemmas have the objective of establishing these requirements. 

Consider two Probabilistic Timed Automata with an identical set of clocks and events: 
$\mathit{PTA}_1=(L_1,\overline{l_1},\chi,\Sigma,I_1,P_1)$ and  
$\mathit{PTA}_2=(L_2,\overline{l_2},\chi,\Sigma,I_2,P_2)$. We assume that the automata have the
same set of urgent events, $\Sigma^u$.
\begin{definition} \label{def:disag}
A state $s$ $\in L_1 \cap L_2$ is a point of disagreement between the two probabilistic timed
automata if 
either they differ on the invariant or they differ in the set of outgoing transitions.
Taking a transition out of a state $s$ as the tuple $(s,z,\sigma,P(2^\chi \times L))$, 
call two transitions different if they disagree on either the guard $z$, or the event label on 
the transition $\sigma$, or the distribution $P(2^\chi \times L)$. 
\end{definition}
The semantic probabilistic systems are $[[\mathit{PTA}_1]]$ and $[[\mathit{PTA}_2]]$ respectively. 
Let $\mathit{States}([[\mathit{PTA}_1]])$ and $\mathit{States}([[\mathit{PTA}_2]])$ 
denote states of the semantic probabilistic systems for $\mathit{PTA}_1$ and 
$\mathit{PTA}_2$ respectively.
The states in the semantic PS are tuples $(s,v)$ where $s$ is a state of the PTA and 
$v$ is a clock valuation. 
\begin{lemma} \label{lem:pta2ps}
A state $(s,v) \in States([[\mathit{PTA}_1]]) \cap States([[\mathit{PTA}_2]])$ is a point of
disagreement (with regard to condition 1 of theorem \ref{thm:ps}) between the two PS implies that
$s$ is a point of disagreement between $\mathit{PTA}_1$ and $\mathit{PTA}_2$. 
\end{lemma}

The condition that labels should also be identical might seem too
restrictive considering that we are only interested in probabilistic reachability.
However, the next set of lemmas will show that when composing PTAs labels are important.

Most real world systems and the 802.11 protocol in particular are modeled as a composition
of PTAs. In a composed system, the above lemma will only tell us whether a particular
common state in the PTA can generate a point of disagreement in the semantic PS. This common state
represents the composed state of all the PTAs composing the model. The next few lemmas 
extend lemma \ref{lem:pta2ps} to the scenario of composed probabilistic timed automata.
\begin{definition} \label{def:difference}
Consider two PTAs formed of compositions, as follows.\\ 
$\mathit{PTA}_1=\mathit{PTA}_1^1 \parallel \mathit{PTA}_2^1 \parallel  \mathit{PTA}_3^1 \parallel ..\parallel \mathit{PTA}_n^1$ and\\ 
$\mathit{PTA}_2=\mathit{PTA}_1^2 \parallel  \mathit{PTA}_2^2 \parallel  \mathit{PTA}_3^2 \parallel .. \parallel \mathit{PTA}_n^2$.\\ 
Define the difference set as the set $D \subseteq \{1,2,..,n\}$ 
such that $\forall i \in D:$ $\mathit{PTA}_i^1 \neq \mathit{PTA}_i^2$ 
and $\forall i \notin D:$ $\mathit{PTA}_i^1 = \mathit{PTA}_i^2$. By equality we 
mean exactly the same automaton in both the compositions (component wise equality of the 
tuples defining them).
\end{definition}
\begin{definition}\label{def:sds}
We define the specific difference set for the index $i \in D$ as 
$D_i \subseteq \mathit{states}(\mathit{PTA}_i^1) \cap \mathit{states}(\mathit{PTA}_i^2)$ where $D_i$ 
is the set of states that disagree across the 
automata as outlined in definition \ref{def:disag}. For every $i \notin D$ set $D_i=\emptyset$.
\end{definition}
\begin{lemma} \label{lem:ptadis}
Consider the composed PTA models of 
Definition \ref{def:difference}. Let $S_{\mathit{common}}$ be the set
of common states between $\mathit{PTA}_1$ and $\mathit{PTA}_2$. A composed 
state in $S_{\mathit{common}}$, say $(l_1,l_2,..,l_n)$ is a point of disagreement 
between $\mathit{PTA}_1$ and $\mathit{PTA}_2$ 
implies that at least one automaton is in its specific difference set.
\end{lemma}
In the composed PTAs of definition \ref{def:difference},  Each state in the semantic PS for a PTA 
is a combination of states and clock valuations of the individual PTA in the composition. The next 
lemma combines lemma \ref{lem:pta2ps} and lemma \ref{lem:ptadis}.
\begin{lemma} [PTA level requirements]\label{lem:ptareq}
A state in  $\mathit{States}([[\mathit{PTA}_1]]) \cap \mathit{States}([[\mathit{PTA}_2]])=
(s_1,s_2..,s_n,v)$ 
is a point of disagreement implies that for at least one $i \in \{1 .. n\}$, the common state $s_i$ of 
both $\mathit{PTA}_i^1$ and $\mathit{PTA}_i^2$ is an element of their specific disagreement set.
\end{lemma}
The purpose of lemma \ref{lem:ptareq} is to identify precisely those states in the \emph{component}
PTA that \emph{may} cause a disagreement in the PS for the composed system.

\subsection{Proof Technique}
We will use the framework in this section to prove the correctness
of our reduced models. Although our objective is the 802.11 protocol, the concept of deterministic
path compression has been developed in a generalized manner anticipating its application to 
other protocols.

To prove that a reduced PTA model ($\mathit{PTA}_2$) corresponding to the original PTA model 
($\mathit{PTA}_1$) is correct, we need to prove that $\mathit{PTA}_1 \stackrel{\mathit{PTA}}{\equiv}_{\phi_1,\phi_2} \mathit{PTA}_2$. Here $\phi_1$ and $\phi_2$ are the corresponding PCTL formulas
in the two models. For our purposes $\phi_1=\phi_2$ since we are interested in proving that
we will arrive at the same result for the same particular PCTL formula. We proceed with the
proof in the following manner.\\
1. Identify the difference set (Definition \ref{def:difference}). Compute the specific 
difference set of each component automaton in the difference set using Definition \ref{def:sds}. 
This is easily done by a visual inspection of the automata.\\
2. Identify composed states where one or more automata are in their specific 
difference set. At this point we use protocol specific proofs to limit such combinations to 
a manageable size. From Lemma \ref{lem:ptadis} we know the set of composed states obtained in this 
step is a superset of the actual difference set across the composed PTA.\\
3. For each composed state, we argue about the possible evolution of the untimed model 
obtained through Definition \ref{pta2psdef}. We show that in each case\\ 
$i)$ There is  the same deterministic dominator in each of 
$[[\mathit{PTA}_1]]$ and $[[\mathit{PTA}_2]]$. This is the hardest part of the proof. However, we use
the fact that the deterministic dominator state in the PS is expressible as the 
combination of a composed state and clock valuation in the PTA. Hence the proofs are in 
terms of the PTA rather than the PS. We generally show that each component 
automaton reaches the state in the composition and progress can only be made when the entire
model is in the composed state.\\
$ii)$ Final states in $[[\mathit{PTA}_1]]$ and $[[\mathit{PTA}_2]]$, corresponding to 
the PCTL formulas $\phi_1$ and $\phi_2$ respectively, are distributed as specified in condition 2 of Theorem \ref{thm:ps}.\\
From Lemma \ref{lem:ptareq} we know that this is sufficient for Theorem \ref{thm:ps} to 
hold. Hence we conclude that at the level of PTAs $\mathit{PTA}_1 \stackrel{\mathit{PTA}}{\equiv}_{\phi_1,\phi_2} \mathit{PTA}_2$.

Deterministic Path Compression, at the level of Probabilistic Systems bears similarity to
weak bisimulation \cite{psim} that can abstract away internal actions. 
However, a notable difference in our approach from 
weak bisimulation is that we are able to change invariants on states in the Probabilistic
Timed Automata. This corresponds to removing time steps (Definition \ref{pta2psdef})
in the corresponding semantic probabilistic system. These time steps are \emph{not} internal
actions because composed probabilistic systems must synchronize on time steps to maintain the
semantics of PTA composition. A possibility would be to apply weak bisimulation to the
final composed model but this would mean fixing the number of stations in the composition.
The reduced models would no longer be valid for the general multi station problem.
\section{Reducing the 802.11 Station Automaton}
\label{realstuff}
For the 802.11 problem, we optimize the station automaton, in multiple steps, starting from the 
original abstract station model of Figure \ref{fig:abstn}.
In each case, the set of final states correspond to the PCTL formula 
$\phi=P_{<\lambda}[\Diamond (\mathit{bc}=k)]$. For every reduction from 
$\mathit{PTA}_1$ to $\mathit{PTA}_2$, we prove the correctness of our optimizations by showing that
$\mathit{PTA}_1 \stackrel{\mathit{PTA}}{\equiv}_{\phi,\phi} \mathit{PTA}_2$. Due to space constraints,
we omit the complete proofs (they are available in \cite{full}) and only motivate the key ideas. 
Our proofs are driven by behavior exhibited by the 802.11 PTA models. 
For example, a key aspect of many of our proofs is the
fact that 802.11 backoff counters are frozen when a busy channel is detected. This sets the 802.11
protocol apart from other contention based protocols such as the 802.3 \cite{ether} and is useful because we 
can essentially ignore stations in backoff when the channel is busy. 
\subsection{Removing the $\mathit{SIFS}$ Wait}
\begin{figure}[h]
\includegraphics[width=8cm]{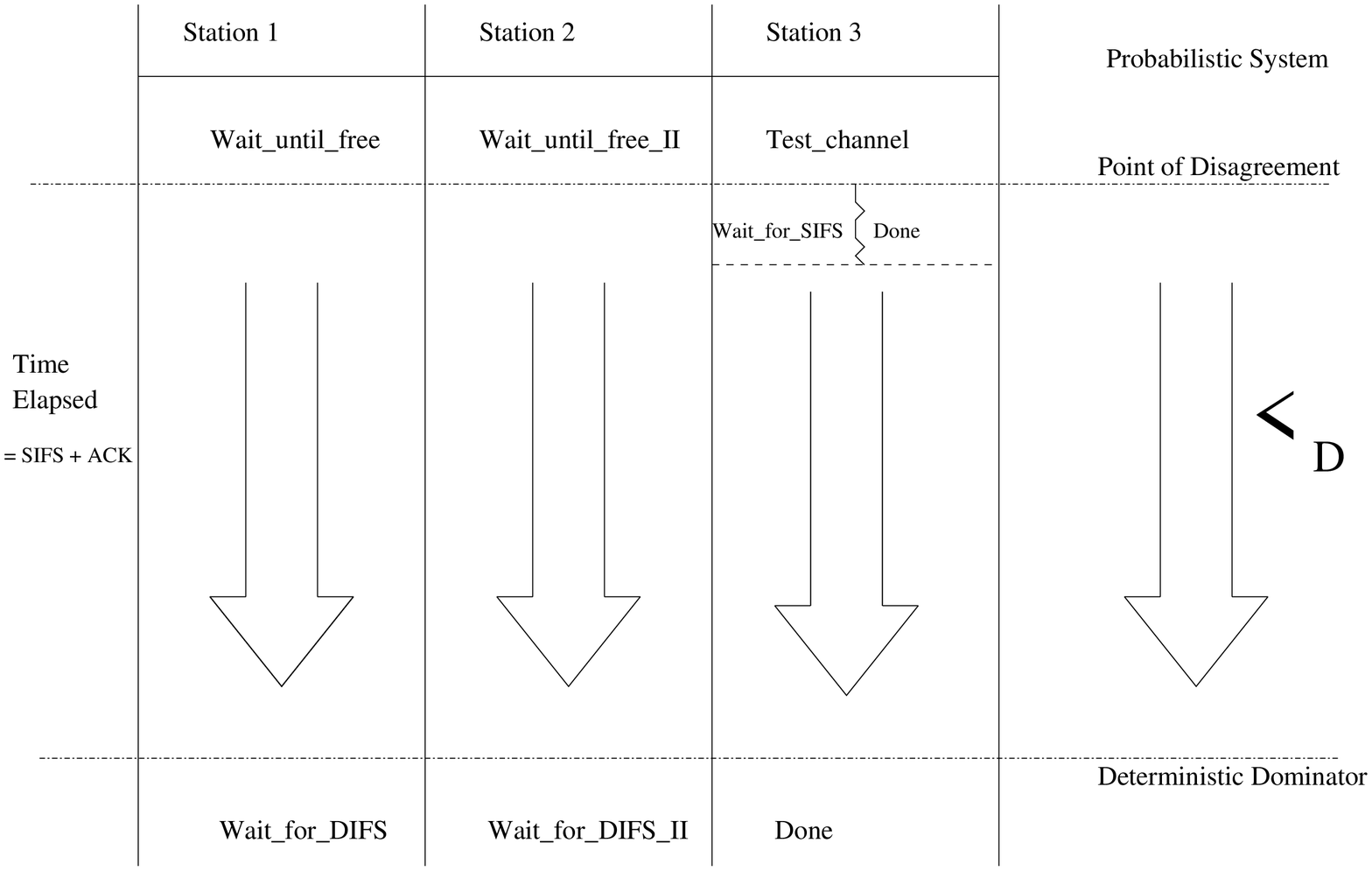}
\caption{Removing the ack protocol - An example with 3 stations}
\label{fig:ack3}
\end{figure}

\begin{figure}[h!]
\centering
\includegraphics[width=8cm,height=10cm]{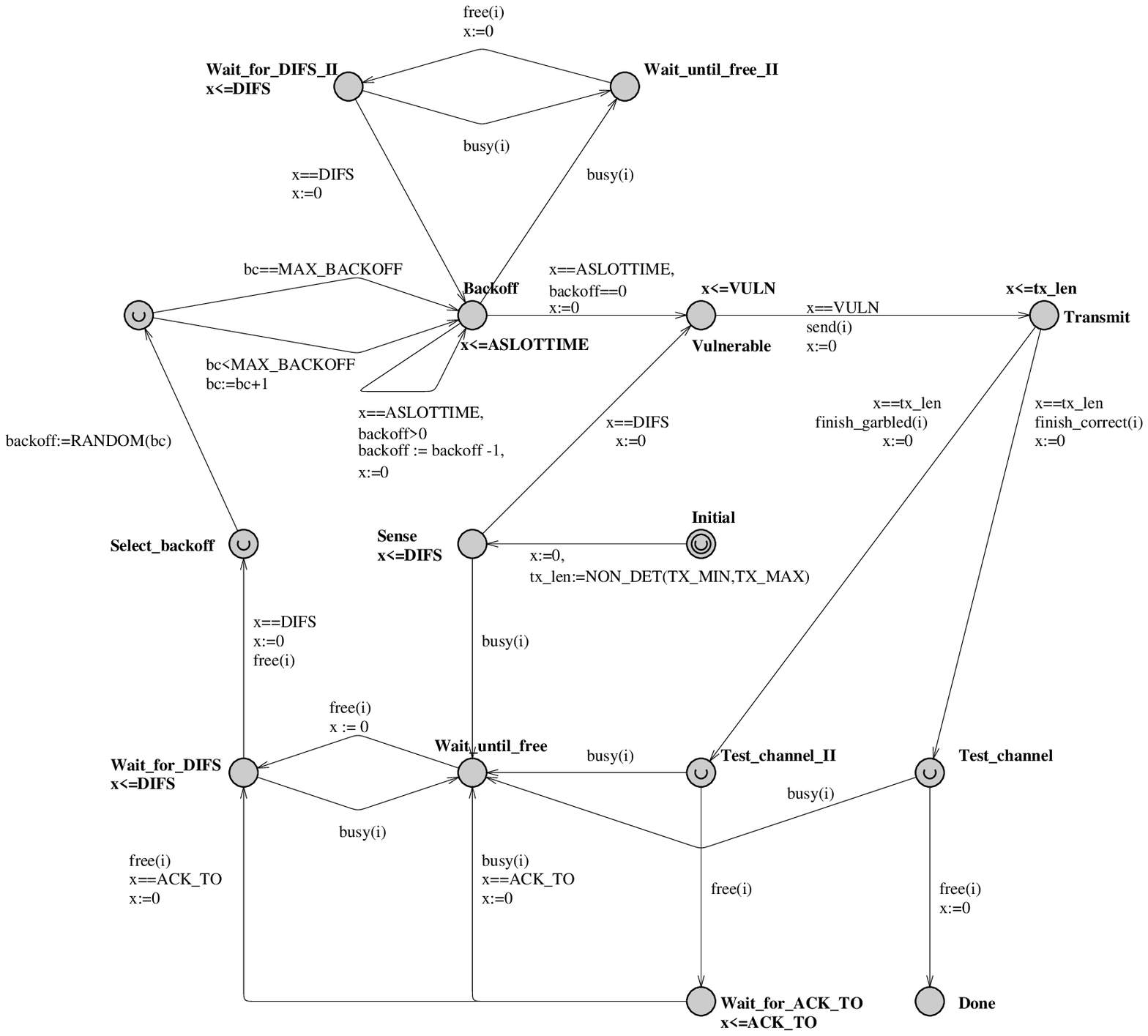}
\caption{PTA model for an Intermediate Abstracted and Reduced Station - The ACK protocol has been removed}
\label{fig:intstn}
\end{figure}

Our first optimization removes the $\mathit{SIFS}$ wait following a successful transmission.
The original model is 
$\mathit{AbsLAN}=\mathit{AbsStn}_1\parallel\mathit{AbsStn}_2\parallel..\parallel\mathit{AbsStn}_n\parallel \mathit{Chan}$ and the reduced model is 
$\mathit{IntLAN}=\mathit{IntStn}_1\parallel\mathit{IntStn}_2\parallel..\parallel\mathit{IntStn}_n\parallel\mathit{Chan}$. The intermediate station model $\mathit{IntStn}$ with the $\mathit{SIFS}$ 
wait removed in shown in Figure \ref{fig:intstn}. The difference set (see Definition 
\ref{def:difference}) includes all the stations and does not include the channel, 
which is unchanged. The specific difference set is only the $Test\_Channel$ 
urgent state immediately after asserting $\mathit{finish\_correct}$. The key idea of the proof is as 
follows: All the other stations will detect the busy
channel and move into the $\mathit{Wait\_until\_free}$ or $\mathit{Wait\_until\_free\_II}$
state. The successfully completing station will move into the $\mathit{Done}$
state while the rest of the stations will move either into $\mathit{Wait\_for\_DIFS}$ or 
$\mathit{Wait\_for\_DIFS\_II}$ 
states, which gives us a deterministic dominator in both the automata ($\mathit{AbsLAN}$ 
and $\mathit{IntLAN}$). In the proof, we exploit the fact that in the 802.11 protocol, 
the backoff counters are frozen when a transmission is detected on the channel. This is modeled 
by the station in \emph{Backoff} moving into the $\mathit{Wait\_until\_free\_II}$ state. The 
key idea of the proof, in an example for three stations, is shown in Figure \ref{fig:ack3}.

\subsection{Removing the DIFS Wait}
\begin{figure}[h]
\includegraphics[width=8cm]{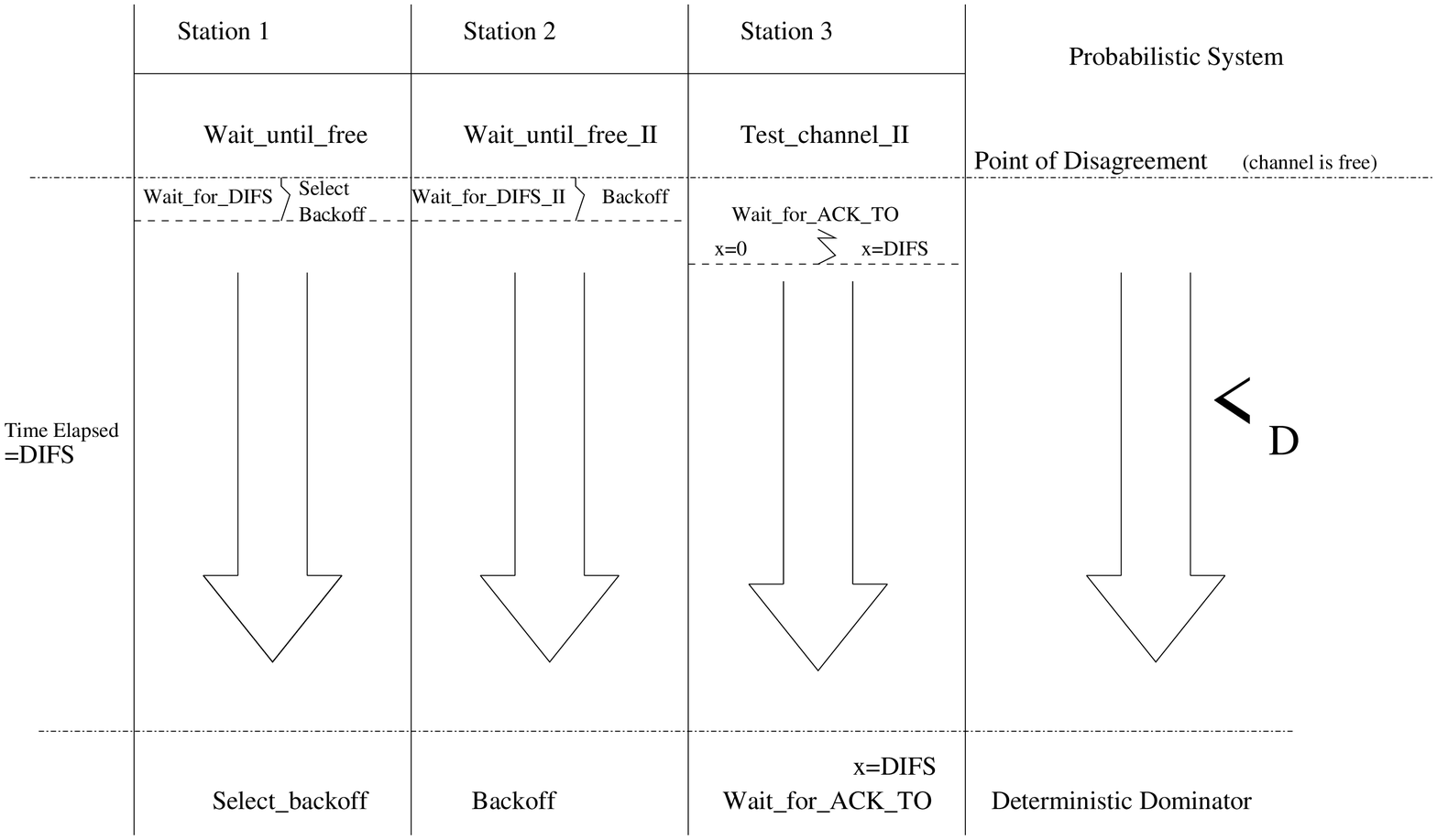}
\caption{Removing the difs wait - An example with 3 stations}
\label{fig:difs3}
\end{figure}

\begin{figure}[h]
\includegraphics[width=8cm,height=10cm]{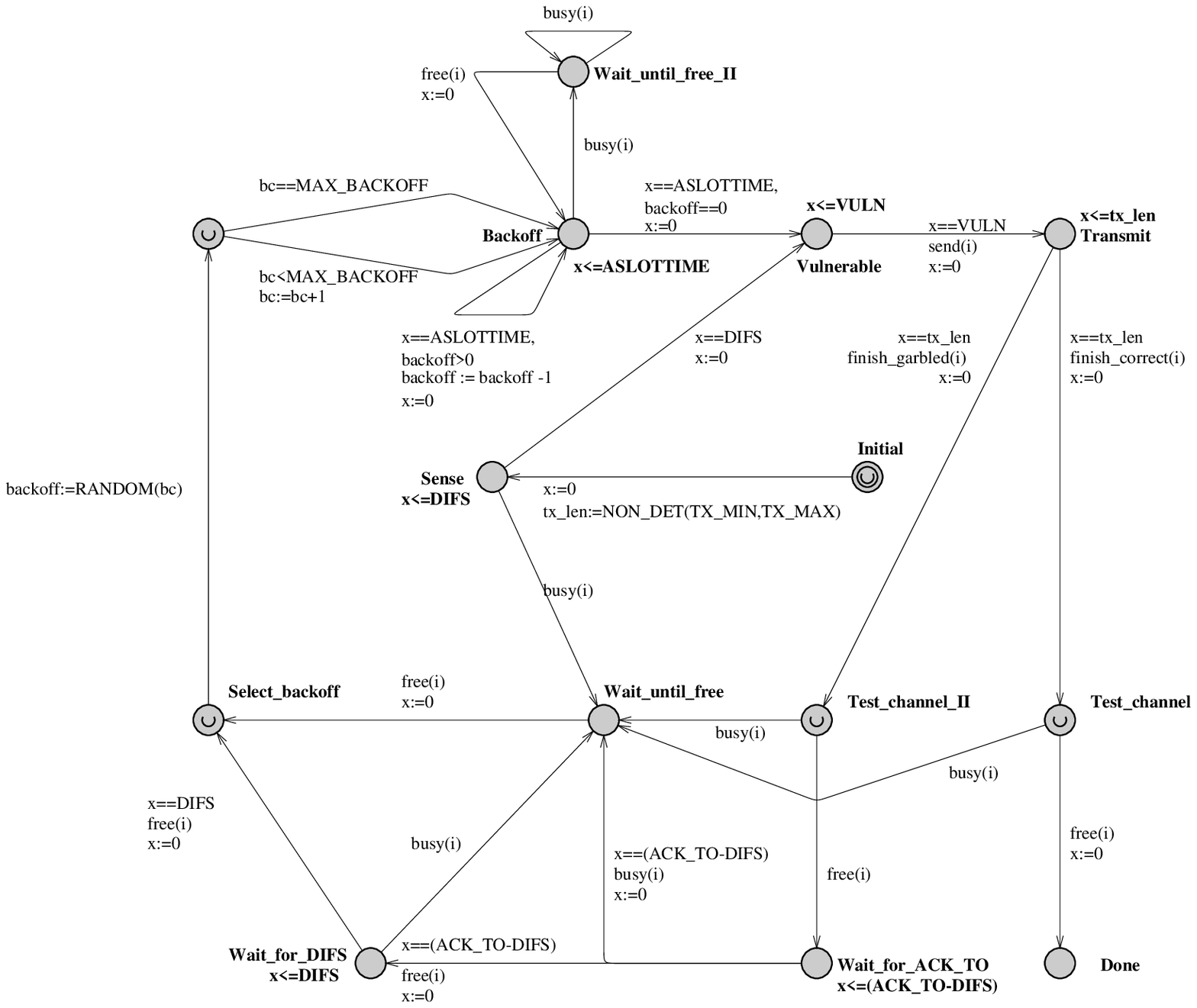}
\caption{PTA model for the Final Abstracted and Reduced Station}
\label{fig:tinystn}
\end{figure}

In the final reduced station model $\mathit{RedStn}$ of Figure \ref{fig:tinystn}, the $\mathit{DIFS}$ 
wait has been removed. The model is given by the composition 
$\mathit{RedLAN}=\mathit{RedStn}_1||\mathit{RedStn}_2..||\mathit{RedStn}_n||\mathit{Chan}$.
Proving the deterministic dominator relationship
is a little more complicated here because we need to consider both collision and successful
transmission cases. In each case however, all stations detect the busy channel and move to
$\mathit{Wait\_until\_free}$ or $\mathit{Wait\_until\_free\_II}$. 
The specific difference set consists of
$\mathit{Wait\_until\_free}$, $\mathit{Wait\_until\_free\_II}$ and $\mathit{Wait\_for\_ACK\_TO}$. 
In the semantic probabilistic
system corresponding to the composed model we can always prove that for any point of disagreement
and for any adversary, there is always a deterministic dominator, which is the
state of the system after the $DIFS$ wait is over. The key idea for a three station example 
is shown in Figure \ref{fig:difs3}.

In $RedStn$ we continue to keep the $Wait\_for\_DIFS$ state. The reason for this is
as follows. It is possible for a station to leave $Wait\_for\_ACK\_TO$ and wait
for $DIFS$ amount of time while all other stations which have not transmitted are sitting in
\emph{Backoff}. Since the amount of time spent in backoff is unpredictable, 
there is no deterministic dominator. Consequently, we cannot simply remove the $DIFS$ wait after 
$Wait\_for\_ACK\_TO$. However,
we may always remove the transition into this state due to the $DIFS$ wait on detecting a
busy channel after transmission.
Again, a key component of the proof is the fact that 802.11 backoff 
counters are \emph{frozen} on detecting a busy channel. This allows us to essentially 
ignore the stations in backoff during transmission.
\subsection{Restricting the allowed transmission length}
The major contributor of state space in the protocol is the large range of allowed 
transmission lengths. The range is from $315 \mu s$ to $15717 \mu s$ and this 
proves to be a significant impediment.

We make a minor change in our PTA models, with the objective of making the 
proofs of equivalence more direct. 
Rather than having a non-deterministic edge 
that selects packet lengths, which are subsequently held constant, we \emph{parameterize} the 
models by a packet length and remove the non-deterministic choice. Hence, we now have
a \emph{series} of PTA models depending on the choice of parameterizations. The allowable assignment 
of packet (transmission) lengths is from 
$\mathit{Par}^{\mathit{full}}$, the set of all possible parameterizations. Each of 
$\mathit{tx\_len}_1, .. ,\mathit{tx\_len}_n$ is assigned a value from the interval $[\mathit{TX\_MIN}, \mathit{TX\_MAX}]$. Formally, 
$\mathit{Par}^{\mathit{full}}= [\mathit{TX\_MIN},\mathit{TX\_MAX}] ^n$.

Consider the reduced set of parameterizations $\mathit{Par}^{\mathit{reduced}} \subset \mathit{Par} ^{\mathit{full}}$ where 
$\mathit{tx\_len}_1=\mathit{TX\_MIN}$ and $\mathit{tx\_len}_{i+1}- \mathit{tx\_len}_i \leq \mathit{VULN}$, $1 \leq i < n$. Here we restrict the
maximum allowable increase in transmission length of one station over its immediate predecessor. 
This eliminates many parameterizations that would have assigned transmission lengths close 
to maximum resulting in a large state space.  We have shown using the 
framework of Section \ref{sect:red} that it is
sufficient to consider only this limited range of transmission lengths. The key objective 
is to show that for every model $\mathit{PTA}_1$ whose parameters are
selected from $\mathit{Par} ^{\mathit{full}}$, there exists a 
model $\mathit{PTA}_2$ whose parameters are contained in $\mathit{Par}^{\mathit{reduced}}$ such that
$\mathit{PTA}_1 \stackrel{\mathit{PTA}}{\equiv}_{\phi,\phi} \mathit{PTA}_2$. Here the
specific difference set is only the $Transmit$ state whose invariant is different in the
two models (due to differing transmission lengths).
Again, we use the fact that 802.11 backoff counters are frozen during transmission. This  means 
that changing the  transmission length has no effect on stations that were in backoff when the channel became busy.
The hardest part is to select a proper model from $\mathit{Par}^{\mathit{reduced}}$ such that any 
m stations in a generalized n-station scenario, that collide by transmitting simultaneously, 
complete transmission in the same order in both the models. This is necessary because an inspection 
of the station automaton shows that during a collision, any station that finishes 
while some other station is still occupying the channel, would detect the busy 
channel and behave differently from the station
that finished last. Hence ensuring that stations finish in the same order leads to the same
deterministic dominator in both the models.
\section{Soft Deadline Verification}
\label{sect:soft}
The probability of meeting soft deadlines, which is the minimum probability of a station
delivering a packet within a certain deadline, is a real time property 
that can be formulated as a probabilistic reachability problem.
For example, in an 802.11 topology of three senders and three receivers, we are interested in
the probability that \emph{every} station successfully transmits its packet
within a given deadline. The reductions presented in this paper, which depend on deterministic path
compression, do not preserve total time elapsed since certain states in the probabilistic timed 
automata where the composite model can count have been removed. As a result, paths are replaced 
with shorter (time wise) versions.

However, one key aspect of our reductions is that they affect deterministic and well-defined segments
of the automata. The intuition is that it should be possible to ``compensate" for the reductions by
using additional available information. For example, removing the acknowledgment protocol has the
effect of subtracting a $\mathit{SIFS} + \mathit{ACK}$ period for every successful transmission made. 
On the other hand removing $\mathit{DIFS}$ wait results in subtracting $\mathit{DIFS}$ from the 
elapsed time for any transmission made.

We begin with the traditional ``decoration" of a PTA in order to
verify real time properties, as exemplified in \cite{fire}. Assume the existence of a composed 
state $\mathit{Done}$, which is the composition of the state $\mathit{Done}$ across 
the components the model. Decorating the PTA involves adding a global clock (say $y$) to the system
that counts total time elapsed and a state $\mathit{Deadline\_exceeded}$. Edges 
are added from each state
other than $\mathit{Done}$, with guard $y \ge \mathit{deadline}$ to  
$\mathit{Deadline\_exceeded}$. Every invariant except at
$\mathit{Done}$ and $\mathit{Deadline\_exceeded}$ is taken in conjunction with 
$y \leq \mathit{deadline}$. The objective is to
model check for the PCTL formula $P_{> \lambda}[\Diamond \mathit{Done}]$, which expresses the
soft deadline property.

We depart from the traditional model by decorating the PTA as follows: We define a non-decreasing
linear function $\phi(y,X)$ on the global clock and numerical system variables (which 
does not include the clock valuation). The global clock $y$ and 
state $\mathit{Deadline\_exceeded}$ are added. 
Edges are added to
$\mathit{Deadline\_exceeded}$ with guard $\phi(y,X) \ge \mathit{deadline}$. 
Each invariant is taken in conjunction
with $\phi(y,X) \leq \mathit{deadline}$. Since the dependence on $X$ may be 
represented as different functions
depending on the current state, we do not depart from the traditional definition of a PTA.
The idea is that while $y$ represents absolute system time, $\phi(y,X)$ represents a
corrected version that takes into account deterministic path compression.

In order to compute real time properties, we annotate the channel with the extra variables
$\mathit{transmissions}$ and $\mathit{successes}$, where each is initialized to zero in the 
start state. The former
is incremented on every synchronization on $\mathit{finish\_correct}$ or 
$\mathit{finish\_garbled}$ while the latter is
incremented only on a synchronization on $\mathit{finish\_correct}$. Their semantics, hence, follow
their nomenclature. In the $\mathit{RedLAN}$ model, without parameter restrictions,
set $\phi(y,X)= y + \mathit{successes} * (\mathit{SIFS} + \mathit{ACK}) + \mathit{transmissions}*\mathit{DIFS}$. This function compensates
for ack protocol removal by adding $\mathit{SIFS}+\mathit{ACK}$ for each successful transmission 
and for $\mathit{DIFS}$ removal
by adding $\mathit{DIFS}$ for every transmission. For the $\mathit{AbsLAN}$ model, we set
$\phi(y,X)= y$, reflecting the standard construction.
Due to space constraints we omit the proof of correctness of our construction here. We essentially
need to repeat the proofs referred to in Section \ref{realstuff}, taking into account
the fact that a clock value assigned to $y$ in the original model will be mapped to
$\phi(y,X)$ in the changed model and we are now model checking for $P_{> \lambda}[\Diamond \mathit{Done}]$.

We intend to extend our technique for retaining soft deadline properties to cover parameter
restrictions in future work.
\section {Verification Results}
\begin{table}
\begin{tabular}{|l|l|l|l|}\hline
Model & States & Transitions & Choices\\\hline
Original & 5958233 & 16563234 & 11437956\\\hline
Optimized & 393958 & 958378 & 598412\\\hline
\end{tabular}
\caption{State space size for two stations - Our optimized model vs. Kwiatkowska et al. \cite{wlan} (original)}
\label{state2}
\end{table}
\begin{table}
\begin{tabular}{|l|l|l|} \hline
Stations&3&4\\\hline
States&1084111823&1377418222475\\\hline  
Transitions&3190610466&5162674182210\\\hline
Choices&1908688031&2958322202754\\\hline
\end{tabular}
\caption{State Space size for three and four stations - Optimized models}
\label{state3}
\end{table}

\begin{table}
\begin{tabular}{|l|l|l|l|}\hline
Backoff&Time&Time&Maximum\\
Counter&original&optimized&probability\\
&(secs)&(secs)&\\\hline 
1&0.69&0.09&1.0\\\hline
2&8.95&1.15&0.18359375\\\hline
3&37.37&6.29&0.01703262\\\hline
4&113.25&29.12&7.9424586e-4\\\hline
5&327.04&120.5&1.8566660e-5\\\hline
6&970.38&508.26&2.1729427e-7\\\hline
\end{tabular}
\caption{Probability of the backoff counter reaching a specified value in the two station case - our model vs. Kwiatkowska et al. \cite{wlan} (original)} 
\label{table:2stn}
\end{table}

\begin{table}
\begin{tabular}{|l|l|l|l|}\hline
Backoff&Iterations&Time&Maximum\\
Counter&&(sec)&Probability\\\hline
1&285&1428&1.0\\\hline
2&107&124&0.59643554\\\hline
3&259&1250&0.10435103\\\hline
4&506&14183&0.008170952\\\hline
5&525&37659&2.83169319e-4\\\hline
6&947&246874&2.85355921e-5\\\hline
\end{tabular}
\caption{Probability of the backoff counter reaching a specified value in the three station case}
\label{table:3stn}
\end{table}

\begin{table}
\centering
\begin{tabular}{|l|l|l|l|}\hline
Protocol&Iterations&Time&Minimum\\
&&(sec)&Probability\\\hline
G.729(1)&85&613&0\\\hline
G.729(2)&256&52388&0.011743453\\\hline
\end{tabular}
\caption{Minimum probability of meeting the soft deadline for the real time case study}
\label{table:realtime}
\end{table}

Our verification platform is a 1.2 GHz Pentium III server with 1.5 GB
of ECC memory and running Linux 2.4. Our experiments used the
Multi-Terminal Binary Decision Diagram (MTBDD) engine of PRISM and all
properties were checked with an accuracy of $10^{-6}$.\\\\

The largest constant in the model, even after 
the optimizations, is $354$. This is still prohibitively large. Hence, before translating into actual 
PRISM models, we perform a time scaling operation \cite{wlan,scaling}.  
For time scaling, we used the backoff 
contention slot length of $50 \mu s$ and divided all guards and invariants by the 
chosen unit, rounding upper bounds on the values of clocks up and lower bounds on the values of 
clocks down. This is the only transformation where we loosen the maximal and minimal 
probabilities to bounds rather than exact values. We also removed the states corresponding to 
acknowledgments from the channel, since we no longer model them.
\subsection{State Space Growth}
The growth in state space for the multi station problem is shown in Tables \ref{state2} and
\ref{state3}. We report the number of states and transitions in the model. We also report 
the number of choices, which is total number of nondeterministic choices summed across
all the states of the model.
In Table \ref{state2} we compare our optimized generalized model for the base
case of two station with the models of \cite{wlan}. We show a significant improvement in
model size. However, when we consider models of three and four stations in Table \ref{state3}, 
the unoptimized models obtained by extending those of \cite{wlan} cannot even be built by 
the model checker within the resources provided. Hence, we only report the state space 
for our own optimized models. 
\subsection{Backoff Counter}
We solve the probabilistic model checking problem of computing the upper bound on the 
probability of the backoff counter on any station reaching a specified value. 

As a starting point, we show that our generalized models are
capable of reproducing the results of the specialized two station
models of \cite{wlan}. In Table \ref{state2} we show state
space cost and in Table \ref{table:2stn} we
show verification costs. Our results are the same as in \cite{wlan} but the verification
costs are lower.

The same results in the case of a three station network is shown in Table \ref{table:3stn}.
The probabilities are higher than the two station case. This is to be expected since
three stations represents more contention for the channel than the two station case.
It has been mentioned that the 3 station problem using the original unoptimized station models are  
beyond the reach of PRISM on our platform.

\subsection{Voice over 802.11: A Real Time Case Study}
An example of soft deadlines for probabilistic verification is given by the following scenario:
An area serviced by a single 100 Mbps 802.3 Local Area Network is occupied by
three overlapping but independent wireless networks, each consisting of an access point 
and $n$ mobile devices. All of these
are equipped with 802.11 capabilities and the access point is distributing
voice data to each of the other $n$ stations in its network.
We consider the specific case where $n=7$ and we use one of two variants of the G.729 \cite{itu}
voice encoding schemes.
In the case of the G.729(1) variant the frame size is 64 bytes and bandwidth requirement is 33.6 Kbps, resulting in a soft deadline of $2196 \mu s$ 
(rounding down to get a stricter integral deadline). On the other hand, in G.729(2)
with a frame rate of 74 bytes and bandwidth requirement of 19.2 Kbps, we have a soft 
deadline $4404 \mu s$. For soft deadline verification, we start with a model paramterized
by the frame size. Subsequently, we use the 
the construction of Section \ref{sect:soft} on $\mathit{RedLAN}$ for verification.

The results for the real time voice delivery problem that translates into 
soft deadlines for a three station topology, are reported in
Table \ref{table:realtime}.  They indicate that in the worst case
G.729(1) cannot meet the soft deadline requirements while G.729(2) has
only a 1\% probability of doing so.
\section{Conclusion}
In this paper, we have introduced deterministic path compression, a
new technique to remove protocol redundancies. We have been successful
in tackling the state space problem for the 802.11 wireless LAN
protocol. We have also shown that it is possible to
compute the minimum probability of meeting soft deadlines in spite of
the optimizations. This is surprising because our optimizations, at
first sight, do not seem amenable to soft deadline verification.

We are yet to reach a solution that can make verifying models with
four or more stations feasible. One option is to use the optimized
models as input to a tool like RAPTURE, which can identify dominators at
the Probabilistic System level, in a manner similar to our approach at
the Probabilistic Timed Automata level.  Our work is still essential
because it brings the model within reach of a tool like RAPTURE.  It
remains to be seen whether significant improvements at the
Probabilistic System level are possible. There are also a number of
extensions to the basic access protocol that we have considered. Modeling these
would justify application of probabilistic verification, which is extremely expensive compared to
simulation, to real world problems.
\paragraph{Acknowledgments:} We thank
Deepak D' Souza for his constructive feedback and Marta Kwiatkowska with 
her colleagues for making the code of the models used in \cite{wlan} available to us.
\bibliographystyle{ieee}
\bibliography{quest}
\end{document}